\documentclass[pra,twocolumn,amsmath,amssymb,10pt,aps,longbibliography,superscriptaddress,citeautoscript,nobibnotes]{revtex4-2}

\usepackage{feynmp}
\usepackage{amsmath}
\usepackage{graphicx}
\usepackage{multirow}
\usepackage{latexsym}
\usepackage{textcomp}
\usepackage{verbatim}
\usepackage{color}
\usepackage{bm}
\usepackage{subfigure}
\usepackage{everysel}
\usepackage{keyval}
\usepackage{ragged2e}
\usepackage{dsfont}
\usepackage{amssymb}
\usepackage{enumitem}
\usepackage{pstricks}
\usepackage{mathtools}

\usepackage{braket}
\usepackage{slashed} 
\usepackage{hyperref}        
\hypersetup{
     colorlinks=true,
     citecolor = red,    
     linkcolor=magenta,
     }

\usepackage{graphicx}
\usepackage{xcolor}
\usepackage{amsmath}
\usepackage{bbold}
\usepackage{amsfonts}
\usepackage{bbm}
\usepackage{comment}
\usepackage{orcidlink}

\newcommand{\osum}{{%
    \setbox0\hbox{\circ}%
    \rlap{\hbox to \wd0{\hss\sum\hss}}\box0
}}

\begin{document}

\title{Enhanced dissipative criticality at an exceptional point}

\author{Jongjun M. Lee\,\orcidlink{0000-0002-9786-1901}}
\thanks{Contact author: jongjun@ualberta.ca}
\affiliation{Department of Physics, University of Alberta, Edmonton, Alberta T6G 2E1, Canada}
\affiliation{Quantum Horizons Alberta \& Theoretical Physics Institute, University of Alberta, Edmonton, Alberta T6G 2E1, Canada}

\begin{abstract}
Exceptional points (EPs) represent non-Hermitian degeneracies where eigenvalues and eigenvectors coalesce, giving rise to enhanced sensitivity and critically damped dynamics. We demonstrate that when an EP coincides with a dissipative phase transition in an extended open Dicke model of two cavities coupled to a collective spin, the critical fluctuations are strongly amplified and governed by modified critical exponents. Numerical results reveal enhanced critical scaling in both the normal and superradiant phases, in agreement with an analytical theory based on the Jordan-block structure of the linearized dynamical matrix at the EP. Our results establish EPs as a mechanism to engineer critical scaling in open quantum systems, with potential applications to critical quantum sensing.
\end{abstract}

\date{\today}
\maketitle

\section{Introduction}
Phase transitions constitute a central paradigm of many-body physics, describing collective phenomena emerging from microscopic interactions~\cite{sondhi1997continuous,sachdev2011quantum}. While equilibrium transitions are governed by the competition between energy and entropy, open quantum systems exhibit qualitatively distinct behavior due to the interplay of coherent and dissipative dynamics~\cite{diehl2008quantum,kessler2012dissipative,carmichael2015breakdown}. In such systems, dissipative phase transitions manifest as nonanalytic changes in the steady state of the Liouvillian, accompanied by the closing of the dissipative gap and divergent fluctuations~\cite{sieberer2016keldysh,minganti2018spectral,rosario2025many}. Rapid experimental progress in cavity and circuit quantum electrodynamics platforms has enabled controlled realizations of driven-dissipative many-body systems~\cite{rodriguez2017probing,fitzpatrick2017observation,fink2018signatures,brookes2021critical}, motivating a deeper understanding of their universal properties~\cite{diehl2010dynamical,marino2016quantum,hwang2018dissipative,belyansky2025phase,sieberer2025universality,lee2026universality}.

Degeneracies of the Liouvillian spectrum constitute a particularly important spectral structure in open quantum dynamics~\cite{khandelwal2021signature,zhou2023accelerating}. Exceptional points (EPs) can arise in effective non-Hermitian Hamiltonians, full Liouvillian superoperators, or linearized Liouvillian operators, and these notions should be carefully distinguished~\cite{minganti2019quantum,minganti2020hybrid,arkhipov2020liouvillian}. At an EP, both eigenvalues and eigenvectors of the corresponding dynamical generator coalesce, rendering the Liouvillian non-diagonalizable and qualitatively modifying the response~\cite{el2018non,arkhipov2020liouvillian}. Such singularities are known to induce enhanced sensitivity~\cite{wiersig2014enhancing,wiersig2016sensors,hodaei2017enhanced,chen2017exceptional,mao2024exceptional} and unconventional dynamical behavior in driven-dissipative systems~\cite{gao2015observation,zhang2018phonon,chen2020revealing,wang2024enhancement}. Yet, whether and how these spectral singularities influence the critical behavior of dissipative phase transitions remains largely unexplored~\cite{hanai2019non,hanai2020critical,zelle2024universal,rahmani2024exceptional,luo2024quantum,cheng2026disorder,nakanishi2026continuous}.

\begin{figure}[t!]
    \centering
    \includegraphics[width=0.95\linewidth]{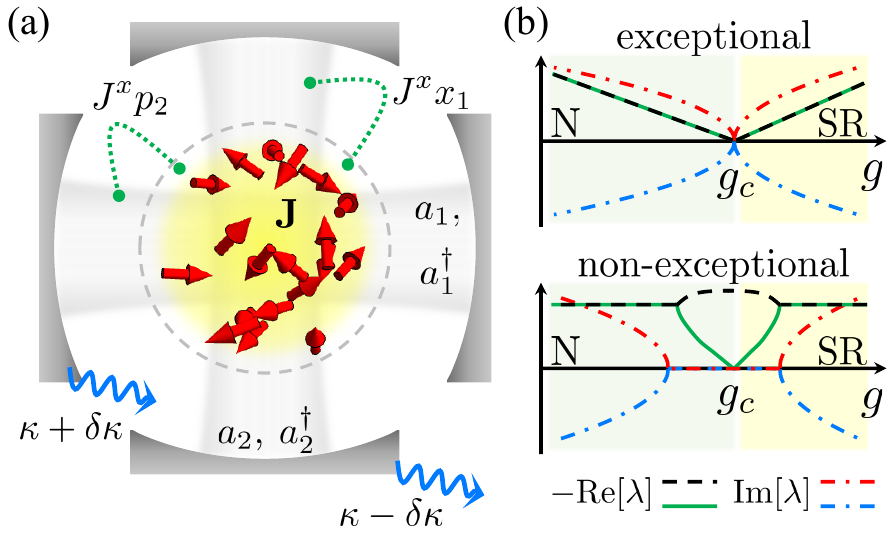}
    \caption{Schematic illustrations of (a) the extended open Dicke model with two cavity modes $a_{1,2}$ coupled to a collective spin $\mathbf{J}$ and (b) the Lindbladian eigenvalues $\lambda$ as a function of the cavity–spin coupling strength $g$. Here, $x_{1}$ and $p_{2}$ denote quadrature operators. Blue wavy arrows indicate single-photon decay of each cavity mode. “(non-)exceptional” denotes the case with (without) the tuning condition in Eq.~(\ref{Eq_tuning}). N and SR label the normal and superradiant phases, respectively.}
    \label{fig1}
\end{figure}

In this work, we investigate how the critical behavior of a dissipative phase transition is reshaped when it coincides with an EP of the linearized Liouvillian~\cite{minganti2019quantum}. We consider an extended open Dicke model comprising two cavity modes coupled to a collective spin, as schematically illustrated in Fig.~\ref{fig1}(a)~\cite{dicke1954coherence,hepp1973equilibrium,kirton2019introduction}. By tuning the cavity detuning and the single-photon loss, we identify a parameter condition in which the Dicke critical point coincides with an EP. We numerically demonstrate a pronounced enhancement of critical fluctuations. For instance, the photon- and magnon-number fluctuations scale with exponent $-2$, in contrast to the conventional exponent $-1$ in the standard Dicke model or away from the tuned condition, both in the normal and superradiant phases~\cite{nagy2011critical,torre2013keldysh,kirton2017suppressing,kirton2019introduction}. The critical exponents are summarized in Table~\ref{Table_summary_1}. Our analytic calculations show that this enhancement arises generically when a dissipative phase transition occurs at an EP.

\section{Dynamics at the exceptional point}
At an EP, eigenvalues and eigenvectors coalesce and the Hamiltonian becomes non-diagonalizable, reducing to a Jordan block~\cite{heiss2012physics,ashida2020non}. As an illustrative example, consider a $2\times2$ case,
\begin{equation}
    \mathbf{S}^{-1} H_{\rm EP} \mathbf{S} =
    \begin{pmatrix}
        \lambda_{0} & 1 \\
        0 & \lambda_{0}
    \end{pmatrix},
\label{Eq_SHS_1}
\end{equation}
where $\lambda_{0}$ is the coalesced eigenvalue and $\mathbf{S}$ is an invertible matrix. The time evolution then acquires a polynomial prefactor,
\begin{equation}
    \psi(t) =  (\mathbb{I}_{2} + \mathbf{C} t) e^{-i\lambda_{0} t} \psi (0),
\label{Eq_psi_1}
\end{equation}
where $\mathbb{I}_{2}$ is the $2\times 2$ identity matrix and $\mathbf{C}$ is a $2\times 2$ matrix that depends on the specific form of the Hamiltonian~\cite{xing2025exceptional}. This linear-in-time factor reflects the Jordan-block structure and is analogous to the behavior of a critically damped oscillator~\cite{LandauLifshitz1976,lelas2023damped}.

In a Lindbladian framework, the Lindbladian becomes non-diagonalizable at an EP~\cite{el2018non,arkhipov2020liouvillian}. In such cases, the density-matrix dynamics can acquire a similar polynomial prefactor, resulting in non-exponential relaxation~\cite{minganti2018spectral}. This modified temporal structure can influence fluctuations and response functions near the EP. Moreover, if an EP coincides with a dissipative critical point, the associated scaling behavior may be affected by the underlying Jordan-block structure.

\section{Extended Dicke model}
We consider an extended open Dicke model consisting of two cavity modes and a collective spin formed by $N$ two-level systems, as shown in Fig.~\ref{fig1}(a). Each cavity interacts with the collective spin and undergoes single-photon loss~\cite{torre2013keldysh,kirton2019introduction}. We consider the thermodynamic limit $N \to \infty$. The density matrix $\rho$ evolves according to the Lindblad master equation $\partial_t \rho = \mathcal{L}[\rho]$, with the Lindbladian superoperator
\begin{equation}
    \mathcal{L}[\rho] = -i[H,\rho]    + (\kappa-\delta\kappa)\mathcal{D}[a_{1}]
    + (\kappa+\delta\kappa)\mathcal{D}[a_{2}],
\label{Eq_Lind_1}
\end{equation}
where $\mathcal{D}[o] = 2 o \rho o^{\dagger} - o^{\dagger} o \rho - \rho o^{\dagger} o$. The Hamiltonian reads
\begin{equation}
H = \Delta (a^{\dagger}_{1}a_{1}-a^{\dagger}_{2}a_{2}) 
+ \omega J^{z} + \frac{2\sqrt{2}g}{\sqrt{N}} J^{x}(x_{1}+p_{2}),
\end{equation}
where $a_j$ is the annihilation operator of the $j$th cavity mode, $\mathbf{J}=\{J^x,J^y,J^z\}$ are collective spin operators with total spin $N/2$, and $\omega>0$ denotes the atomic transition frequency. The quadratures are defined as
\begin{equation}
x_j = \frac{a_j + a_j^{\dagger}}{\sqrt{2}}, \: p_j = \frac{a_j - a_j^{\dagger}}{i\sqrt{2}}.
\label{Eq_quad_1}
\end{equation}
We denote the cavity-spin coupling strength by $g \ge 0$ and $\kappa \pm \delta\kappa > 0$ are the photon loss rates of the two cavities. We denote expectation values as $\langle \mathcal{O} \rangle = \mathrm{Tr}[\mathcal{O}\rho_{\rm ss}]$ where $\rho_{\rm ss}$ is the steady state. Compared to the conventional Dicke model, the present setup includes a second cavity mode~\cite{dicke1954coherence,hepp1973equilibrium}.

Our Lindbladian possesses a weak parity symmetry generated by $\Pi=\exp[i\pi(\sum_{j}a^{\dagger}_{j}a_{j}+J^{z})]$. This operator commutes with the Hamiltonian, $[\Pi,H]=0$, and anticommutes with the dissipators, $\{ \Pi,a_j \}=0$. As a consequence, the full Liouvillian is invariant under the parity transformation, $\mathcal{L}[\Pi\rho\Pi^{\dagger}]=\Pi \mathcal{L}[\rho]\Pi^{\dagger}$, independently of the system parameters~\cite{buvca2012note,albert2014symmetries,lieu2020symmetry}. Nevertheless, this symmetry need not be reflected in the steady state, since it can be spontaneously broken in the thermodynamic limit~\cite{minganti2018spectral}.

We regard the present model as an effective open Dicke model rather than as a microscopic model for a specific cavity-QED implementation~\cite{kirton2019introduction}. For a direct implementation based on dipole coupling, the $A^{2}$ term must additionally be considered and may modify or suppress the superradiant phase transition~\cite{bialynicki1979no,nataf2010no,rokaj2018light,garziano2020gauge}.

\section{Mean-field analysis}
At the mean-field level, the equations of motion for the first moments yield two steady-state solutions~\cite{EM_ref}. We define $\alpha_j=\langle a_j\rangle$ and $S^i=\langle J^i\rangle$. The trivial (normal) solution is $\alpha_{1,2}=0, \: S^{x,y}=0, \: S^z=-\frac{N}{2},$ which preserves the parity symmetry. The nontrivial (superradiant) solution with photon condensation reads
\begin{equation}
\begin{aligned}
\alpha_{1} &= \frac{2gS^{x}}{\sqrt{N}} 
\frac{-\Delta-i(\kappa+\delta\kappa)}
{\Delta^{2}+\kappa^{2}-\delta\kappa^{2}+2i\Delta \delta\kappa},\\
\alpha_{2} &= \frac{2gS^{x}}{\sqrt{N}} 
\frac{i\Delta+(\kappa-\delta\kappa)}
{\Delta^{2}+\kappa^{2}-\delta\kappa^{2}+2i\Delta \delta\kappa}
\end{aligned}
\end{equation}
for the cavity fields, and
\begin{equation}
S^{x} = \pm \sqrt{\frac{N^{2}}{4}-(S^{z})^{2}},\:
S^{y}=0,\:
S^{z} = - \frac{\omega N D}{32g^{2}\Delta\kappa\delta\kappa}
\end{equation}
for the collective spin, where
$D = 4\Delta^{2}\delta\kappa^{2} +(\Delta^{2}+\kappa^{2}-\delta\kappa^{2})^{2}$. This solution breaks the weak parity symmetry~\cite{torre2013keldysh,kirton2019introduction}, and the two possible signs of $S^x$ correspond to spontaneous symmetry breaking of this parity. At the critical coupling
\begin{equation}
g_{c} = \frac{1}{4}
\sqrt{\frac{\omega D}{\Delta\kappa\delta\kappa}},
\label{Eq_gc}
\end{equation}
the nontrivial solution continuously merges with the trivial one, $S^z = -N/2$ and $S^x = 0$. Thus, $g_c$ marks the onset of the superradiant phase transition, as we show below. For small deviations from $g_{c}$, the condensates $\alpha_j$ and the transverse spin component $S^x$ scale as $\sqrt{g-g_c}$, as in the conventional Dicke model~\cite{nagy2011critical,torre2013keldysh,kirton2017suppressing,kirton2019introduction}.

\begin{table}[t!]
\setlength{\tabcolsep}{6pt}
\begin{tabular}{lccccc}
\hline \hline 
& $\Gamma_{\rm ADR}$ & $\delta n_{1}$ & $\delta n_{2}$ & $\delta n_{b}$ & $\mu$ \\ \hline
Dicke & $1$ & $-1$ & - & $-1$ & $1/2$ \\ \hline
Exceptional & $1$ & $-2$ & $-2$ & $-2$ & $3/2$ \\ \hline
Non-exceptional & $1$ & $-1$ & $-1$ & $-1$ & $1/2$ \\ \hline
\end{tabular}
\caption{Summary of the critical exponents defined by $X \propto |g-g_{c}|^{\nu_X}$. ``Dicke” denotes the conventional open Dicke model, while ``Exceptional" and ``Non-exceptional" refer to the extended model at and away from the exceptional-point condition, respectively. $\Gamma_{\rm ADR}$, $\delta n_{1,2}$, $\delta n_{b}$, and $\mu$ represent the asymptotic decay rate, cavity photon-number fluctuations, magnon-number fluctuation, and purity.}
\label{Table_summary_1}
\end{table}

\section{Gaussian fluctuation theory}
Expanding the master equation around the mean-field solution and linearizing the fluctuations, we obtain the Langevin equations for the quadrature fluctuations~\cite{EM_ref},
\begin{equation}
    \frac{d \mathbf{v}(t)}{dt} = \mathbf{A} \mathbf{v}(t) + \bm{\xi}(t),
\label{Eq_Langevin_1}
\end{equation}
where $\mathbf{v} = (\delta x_{1},\: \delta p_{1},\: \delta x_{2},\: \delta p_{2},\: x_{b},\: p_{b})^{\rm T}$ and the noise satisfies $\langle \bm{\xi}(t)\bm{\xi}^{\rm T}(t')\rangle=\mathbf{D}\delta(t-t')$. Here $\delta x_{j}$ and $\delta p_{j}$ denote the quadrature fluctuations of the cavity fields, while $x_b$ and $p_b$ are the quadratures of the spin fluctuation described by a magnon operator $b$~\cite{EM_ref}. The drift matrix reads
\begin{equation}
\begin{aligned}
\mathbf{A}=
\begin{pmatrix}
-\kappa_{-} & \Delta & 0 & 0 & 0 & 0 \\
-\Delta & -\kappa_{-} & 0 & 0 & -2\tilde{g} & 0 \\
0 & 0 & -\kappa_{+} & -\Delta & +2\tilde{g} & 0 \\
0 & 0 & \Delta & -\kappa_{+} & 0 & 0 \\
0 & 0 & 0 & 0 & 0 & \Omega \\
-2\tilde{g} & 0 & 0 & -2\tilde{g} & -\Omega & 0
\end{pmatrix},
\end{aligned}
\end{equation}
and the diffusion matrix is
\begin{equation}
\mathbf{D}=\mathrm{diag}(\kappa_{-},\:\kappa_{-},\: \kappa_{+},\:\kappa_{+},\: 0,\: 0),
\end{equation}
where $\Omega=\omega\cos\theta-(4g\sin\theta/\sqrt{N})(\mathrm{Re}[\alpha_{1}]+\mathrm{Im}[\alpha_{2}])$, $\tilde g=g\cos\theta$, and $\kappa_{\pm}=\kappa\pm\delta\kappa$, with $\theta$ the angle of the mean-field spin~\cite{EM_ref}. The matrix $\mathbf{A}$ governs the linearized Liouvillian dynamics around the mean-field solution. In the normal phase ($\alpha_j=0$, $\theta=0$), one finds $\text{Det}[\mathbf{A}]=0$ at $g=g_c$ [Eq.~(\ref{Eq_gc})], indicating the closing of the dissipative gap at the critical point.

\section{Exceptional point and decay}
By tuning the cavity detuning $\Delta$ and the single-photon losses $\kappa\pm \delta\kappa$ such that
\begin{equation}
    \Delta = \sqrt{\kappa^{2}-\delta\kappa^{2}+2\kappa\sqrt{\kappa^{2}-\delta\kappa^{2}}},
\label{Eq_tuning}
\end{equation}
the drift matrix $\mathbf{A}$ becomes singular precisely at the critical coupling $g_{c}$. At $g=g_{c}$, the zero eigenvalue has algebraic multiplicity two, while
\begin{equation}
    \dim \ker \mathbf{A} = 1,
\end{equation}
so that the geometric multiplicity is strictly smaller than the algebraic multiplicity. Consequently, $\mathbf{A}$ is non-diagonalizable at criticality, and the superradiant phase transition coincides with an EP of the drift matrix governing the linearized Liouvillian dynamics~\cite{minganti2019quantum}. We refer to systems that satisfy (do not satisfy) the tuning condition in Eq.~(\ref{Eq_tuning}) as exceptional (non-exceptional) cases.

Solving the eigenvalue problem of $\mathbf{A}$ near $g=g_c$, we find the two slowest decaying modes~\cite{suppl_ref}. Their eigenvalues exhibit nearly identical real parts and imaginary parts of equal magnitude with opposite signs, as shown in Fig.~\ref{fig1}(b).
\begin{equation}
    \mathrm{Re}[\lambda] \propto -|g-g_{c}|, \: \mathrm{Im}[\lambda] \propto \pm \sqrt{|g-g_{c}|}.
\label{Eq_Eval_1}
\end{equation}
Accordingly, the two critical modes share an asymptotic decay rate that vanishes linearly, $\Gamma_{\rm ADR}\propto |g-g_{c}|$, as in the conventional Dicke model~\cite{nagy2011critical,torre2013keldysh,kirton2017suppressing,kirton2019introduction}. This linear closing of $\Gamma_{\rm ADR}$ persists even away from the exceptional condition in Eq.~(\ref{Eq_tuning})~\cite{suppl_ref}. However, when the critical point coincides with the EP, the Jordan-block structure [Eq.~(\ref{Eq_psi_1})] alters the temporal evolution, suggesting that static critical scaling may also be modified despite the identical decay-rate behavior.

\section{Static fluctuations}
We compute the static Gaussian fluctuations from the linearized Langevin equation [Eq.~(\ref{Eq_Langevin_1})]. The steady-state covariance matrix $\mathbf{V}$ satisfies the Lyapunov equation
\begin{equation}
\mathbf{A}\mathbf{V}+\mathbf{V}\mathbf{A}^{\rm T}+\mathbf{D}=0,
\end{equation}
where $(\mathbf{V})_{ij}=\langle (\mathbf{v})_{i}(\mathbf{v})_{j}\rangle_{s}$ and $\langle \cdot \rangle_s$ is the symmetrized expectation value~\cite{suppl_ref}. For each parameter set, we numerically evaluate $\mathbf{V}$ near the critical point $g_c$.

\begin{figure}
    \centering
    \includegraphics[width=1.0\linewidth]{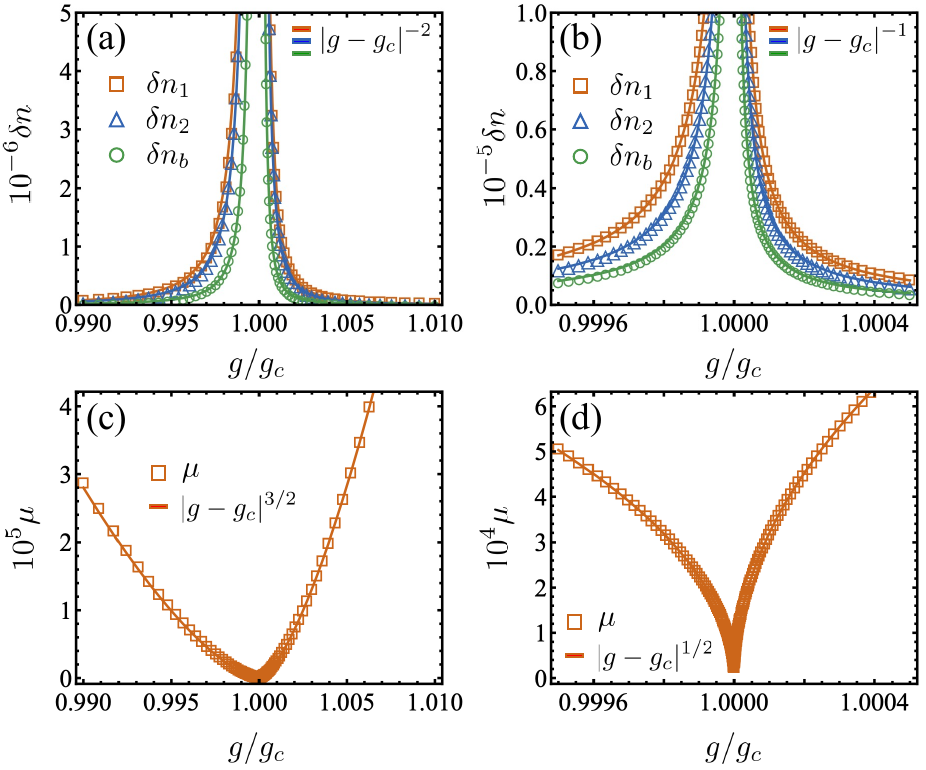}
    \caption{Static Gaussian observables as functions of $g$ near the critical point $g_c$. Symbols represent numerical data, and solid lines denote power-law fits. (a),(b) Number fluctuations $\delta n_j$ for the exceptional and non-exceptional cases, respectively, where $j=1,2,b$ labels the two cavity modes and the magnon mode. (c),(d) Purity $\mu$ for the exceptional and non-exceptional cases, respectively.}
    \label{fig2}
\end{figure}

Figures~\ref{fig2}(a) and (b) show the number fluctuations as functions of $g$ for the exceptional and non-exceptional cases, respectively. The number fluctuation is defined as
\begin{equation}
\delta n_{j}= \frac{1}{2}(\langle \delta x_{j}^{2}\rangle+ \langle \delta p_{j}^{2}\rangle-1),
\end{equation}
where $j=1,2,b$ labels the two cavity modes and the magnon mode. Near criticality, the two cases exhibit clearly distinct scaling behaviors. In the non-exceptional case, the fluctuations diverge as 
\begin{equation}
\delta n_j \propto |g-g_c|^{-1},    
\end{equation}
recovering the conventional Dicke critical exponent~\cite{nagy2011critical,torre2013keldysh,kirton2017suppressing,kirton2019introduction}. By contrast, when the dissipative critical point coincides with the EP, the fluctuations display a much steeper divergence, 
\begin{equation}
\delta n_j \propto |g-g_c|^{-2}.    
\end{equation}
Thus, although the asymptotic decay rate $\Gamma_{\rm ADR}$ vanishes linearly in both cases, the coincidence with the EP enhances Gaussian fluctuations and modifies their critical exponent.

A similar contrast appears in the steady-state purity $\mu = \text{Tr}[\rho^{2}_{\rm ss}]$, shown in Figs.~\ref{fig2}(c) and (d). While the purity approaches zero at $g_c$ in both cases~\cite{suppl_ref}, their critical profiles differ qualitatively. In the non-exceptional case, the purity follows the conventional square-root scaling, 
\begin{equation}
\mu \propto |g-g_c|^{1/2},   
\end{equation}
leading to a sharp cusp at criticality~\cite{nagy2011critical,torre2013keldysh,kirton2017suppressing,kirton2019introduction}. In contrast, under the exceptional condition, the purity vanishes as 
\begin{equation}
\mu \propto |g-g_c|^{3/2},    
\end{equation}
exhibiting a flatter curvature near $g_c$. The transition from square-root to three-halves scaling reflects a qualitative reshaping of the steady-state landscape, indicating that the non-diagonalizable Jordan-block structure at the EP directly imprints on static observables.

\section{Origin of enhanced scaling}
The enhanced critical scaling can be understood analytically from the Lyapunov equation~\cite{mari2012cooling,woolley2014two}. Using the Langevin dynamics [Eq.~(\ref{Eq_Langevin_1})], the steady-state covariance matrix admits the formal solution
\begin{equation}
    \mathbf{V} = \int_{0}^{\infty} du \, e^{\mathbf{A}u}\mathbf{D}e^{\mathbf{A}^{\rm T}u}.
\label{Eq_V_1}
\end{equation}
Near the critical point, the drift matrix can be brought into its Jordan form, $\mathbf{A}=\mathbf{S}\mathbf{J}\mathbf{S}^{-1}$, where $\mathbf{S}$ contains eigenvectors and a generalized eigenvector. The Jordan matrix reads
\begin{equation}
\mathbf{J}=\mathrm{diag}(\lambda_{+},\lambda_{-},\lambda_{2},\lambda_{3},\lambda_{4},\lambda_{5})+\mathbf{E}_{12},
\end{equation}
with $(\mathbf{E}_{12})_{ij}=\delta_{i,1}\delta_{j,2}$. For the two slowest modes we take
\begin{equation}
    \lambda_{+}=-\epsilon+i c_{2}\sqrt{\epsilon},\: \lambda_{-}=-c_{1}\epsilon-i c_{2}\sqrt{\epsilon},
\end{equation}
where $\epsilon\propto |g-g_c|$ and $c_{1,2}$ is real, consistent with Eq.~(\ref{Eq_Eval_1})~\cite{suppl_ref}. At $\epsilon=0$, the two eigenvalues coalesce and $\mathbf{A}$ becomes non-diagonalizable, signaling the EP. Using
\begin{equation}
\mathbf{J}^{n}=\mathrm{diag}(\lambda_{+}^{n},\lambda_{-}^{n},\ldots)+\frac{\lambda_{+}^{n}-\lambda_{-}^{n}}{\lambda_{+}-\lambda_{-}}\mathbf{E}_{12},
\end{equation}
the matrix exponential can be written as
\begin{equation}
e^{\mathbf{A}u}=\mathbf{S}\left(\bm{\Omega}_{\lambda}+f^{\lambda_{\pm}}_{u}\mathbf{E}_{12}
\right)\mathbf{S}^{-1},
\end{equation}
where $\bm{\Omega}_{\lambda}=\text{diag}(e^{\lambda_{+}u},\dots,e^{\lambda_{5}u})$, and $f^{\lambda_{\pm}}_{u}=\frac{e^{\lambda_{+}u}-e^{\lambda_{-}u}}{\lambda_{+}-\lambda_{-}}$. Substituting into the integral expression for $\mathbf{V}$ gives
\begin{equation}
\mathbf{V}=\int_{0}^{\infty} du \, \mathbf{S}(\bm{\Omega}_{\lambda}+f^{\lambda_{\pm}}_{u}\mathbf{E}_{12})
\tilde{\mathbf{D}}(\bm{\Omega}_{\lambda}+f^{\lambda_{\pm}}_{u}\mathbf{E}_{12}^{\rm T})
\mathbf{S}^{\rm T},
\label{Eq_V_2}
\end{equation}
where $\tilde{\mathbf{D}}=\mathbf{S}^{-1}\mathbf{D}(\mathbf{S}^{-1})^{\rm T}$.

The integrals involving $\bm{\Omega}_{\lambda}$ and powers of $f^{\lambda_{\pm}}_{u}\mathbf{E}_{12}$ generate distinct powers of $|g-g_{c}|$ in Eq.~(\ref{Eq_V_2}). In particular, the scaling $|g-g_{c}|^{-2}$ is the most relevant contribution in the covariance matrix. This explains the enhanced critical exponent of the number fluctuations. Importantly, while the asymptotic decay rate $\Gamma_{\rm ADR}\propto |g-g_{c}|$ remains linear as in the conventional Dicke model, the Jordan-block structure produces higher-order divergences in static observables~\cite{nagy2011critical,torre2013keldysh,kirton2019introduction}. The enhancement, therefore, originates from the singularity of the linearized Liouvillian operator and is not a model-specific feature, but a general consequence of an EP coinciding with a dissipative critical point.

\begin{figure*}
    \centering
    \includegraphics[width=0.9\linewidth]{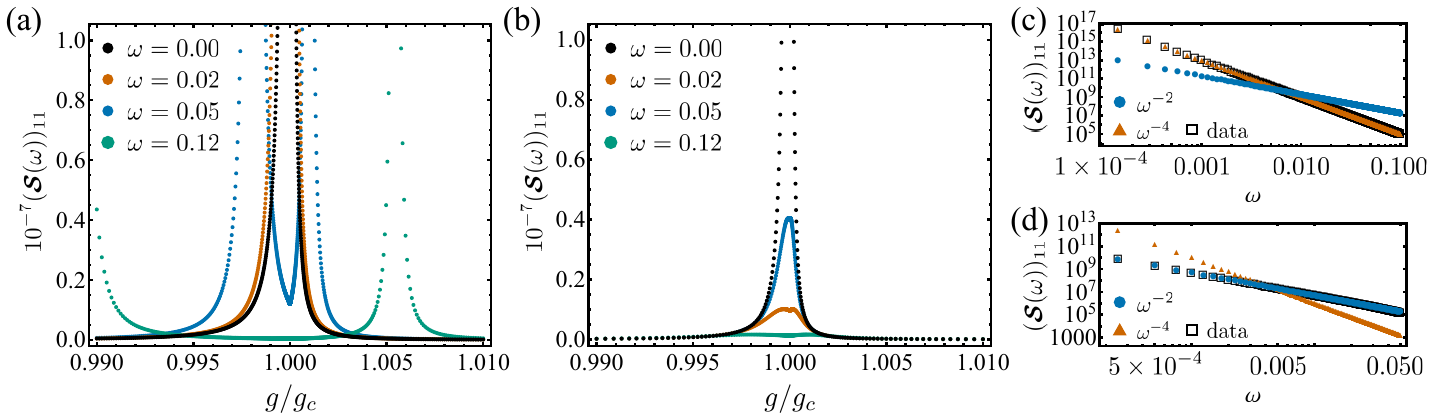}
    \caption{Noise spectrum near and at the critical point. (a,b) Noise spectrum $(\bm{\mathcal{S}}(\omega))_{11}$ as a function of the coupling $g$ around $g_{c}$ for several values of $\omega$, indicated in the panels. (c,d) Noise spectrum $(\bm{\mathcal{S}}(\omega))_{11}$ at the critical point $g=g_{c}$ as a function of $\omega$. Panels (a,c) correspond to the exceptional case, where the dissipative phase transition coincides with the EP, while (b,d) show the non-exceptional case.}
    \label{fig3}
\end{figure*}

\section{Noise spectrum}
We next investigate the symmetrized noise spectrum, which characterizes the frequency dependence of second-moment fluctuations~\cite{clerk2010introduction}. It is defined as
\begin{equation}
    \bm{\mathcal{S}}(\omega)= \mathbf{R}(\omega)\mathbf{D}\mathbf{R}^{\dagger}(\omega),
\end{equation}
where $\mathbf{R}(\omega)=(i\omega\mathcal{I}-\mathbf{A})^{-1}$ and $\mathcal{I}$ is the identity matrix. The frequency dependence of $\bm{\mathcal{S}}(\omega)$ differs qualitatively between the exceptional and non-exceptional cases, particularly at the critical point $g_c$. 

In the non-exceptional case, $\mathbf{R}(\omega)$ is diagonalizable, and since the slowest eigenvalue of the drift matrix vanishes at criticality, one obtains $\mathbf{R}(\omega)\propto \omega^{-1}$. Consequently, the noise spectrum scales as $\bm{\mathcal{S}}(\omega)\propto \omega^{-2}$. In contrast, when the EP condition is satisfied, $\mathbf{R}(\omega)$ becomes non-diagonalizable at $g_c$. For instance, in a $2\times2$ Jordan block,
\begin{equation}
\mathbf{S}_{0}^{-1}\mathbf{R}(\omega)\mathbf{S}_{0}
=
\begin{pmatrix}
i\omega & -1 \\ 0 & i\omega
\end{pmatrix}^{-1}
=
\begin{pmatrix}
\frac{1}{i\omega} & \left(\frac{1}{i\omega}\right)^2 \\ 0 & \frac{1}{i\omega}
\end{pmatrix},
\end{equation}
with an invertible matrix $\mathbf{S}_{0}$. The additional polynomial term leads to $\bm{\mathcal{S}}(\omega)\propto \omega^{-4}$, in contrast to the non-exceptional case.

We numerically evaluate the noise spectrum for both parameter regimes and plot the results of $(\bm{\mathcal{S}}(\omega))_{11}$ in Fig.~\ref{fig3}. In the exceptional case, the spectrum exhibits two sharp peaks at finite $\omega$ as the coupling $g$ approaches criticality. By contrast, the non-exceptional case shows a single broadened peak centered at $g_c$, with finite $\omega$ smoothing the divergence. The peak splitting originates from the remaining imaginary component of the drift-matrix eigenvalues of the slowest mode, while the sharpness of the peaks reflects the singular linearized Liouvillian structure associated with the EP~\cite{torre2013keldysh}. At $g=g_c$, the frequency dependence confirms the predicted scaling. The exceptional case follows $\omega^{-4}$, whereas the non-exceptional case follows $\omega^{-2}$, consistent with the Jordan-block structure at the EP.

\section{Further enhancement}
Higher-order EPs may provide a route to further enhance critical scaling~\cite{hodaei2017enhanced,mandal2021symmetry}. In the present case, a second-order EP produces a $2\times2$ Jordan block [Eq.~(\ref{Eq_SHS_1})] and a temporal structure [Eq.~(\ref{Eq_psi_1})], which leads to the enhanced static divergence discussed above. More generally, if three slow modes coalesce at criticality to form a third-order EP (EP-3), the wavefunction acquires a quadratic prefactor $\psi(t)= e^{-i\lambda_{0} t}(\mathbb{I}_{3}+\mathbf{C}_1 t+\mathbf{C}_2 t^2)\psi(0)$, corresponding to a $3\times3$ Jordan block~\cite{schnabel2017pt,zhang2019higher,kim2023third}. In the Lyapunov integral [Eq.~(\ref{Eq_V_2})], higher-order polynomial prefactors generate stronger power-law divergences (see Appendix~\ref{Appx_EP3}), so that increasing the Jordan-block order systematically enhances the scaling of static observables.

In our model, the imaginary part of the drift-matrix eigenvalues vanishes at the critical point~\cite{suppl_ref}. However, this is neither a necessary condition for a dissipative phase transition nor for the coincidence between the critical point and an EP~\cite{minganti2018spectral}. In general, the imaginary part can become zero away from criticality, so that near $g_c$ the degenerate drift-matrix eigenvalues can be purely real, e.g., $\lambda \propto -|g-g_c|$~\cite{morrison2008dissipation,hwang2018dissipative,lyu2024multicritical,lee2026universality}. This situation is, in fact, more conventional, as both the standard Dicke model and the non-exceptional regime of our extended model exhibit a real slow mode near criticality~\cite{nagy2011critical,torre2013keldysh,kirton2017suppressing,kirton2019introduction}. If degenerate modes with such real linear closing were to coincide with an EP at the critical point, the Lyapunov analysis of Eq.~(\ref{Eq_V_2}) indicates that the most divergent contribution to the covariance matrix would scale as $|g-g_c|^{-3}$. This provides an alternative route to further enhancing critical behavior.

\section{Conclusion and outlook}
We have shown that when a dissipative phase transition coincides with an EP, its static critical behavior is qualitatively reshaped. In an extended open Dicke model, a specific tuning of the cavity detuning and photon losses causes the superradiant critical point to merge with a Liouvillian EP. While the asymptotic decay rate closes with exponent $1$, identical to the conventional Dicke transition, the second-moment fluctuations and the purity exhibit enhanced critical scaling. In particular, the critical exponent of the number fluctuations changes from $-1$ to $-2$, and that of the purity from $1/2$ to $3/2$, as summarized in Table~\ref{Table_summary_1}.

This behavior originates from the non-diagonalizable Jordan-block structure of the linearized Liouvillian at criticality, indicating that the enhanced scaling arises from the spectral singularity itself rather than from changes in the gap closing. More broadly, since critical fluctuations play a central role in quantum-enhanced sensing schemes, our results suggest that EP criticality may provide a new route toward critical quantum sensing and metrological amplification in driven-dissipative systems.

Future work may explore how exceptional points modify the anomalous critical scaling recently proposed in frustrated Dicke models~\cite{zhao2022frustrated,zhao2023anomalous,lee2026frustrated}. It would also be interesting to investigate dissipative phase transitions in systems with strong symmetries, where the steady-state structure differs qualitatively from that of the weak-symmetric model considered here~\cite{lee2026universality,buvca2012note,albert2014symmetries}. Another important direction is to determine how the coincidence of a dissipative critical point and an exceptional point modifies finite-size scaling beyond the thermodynamic-limit Gaussian theory considered in this work~\cite{carmichael2015breakdown,curtis2021critical}.

\acknowledgements
We thank Igor Boettcher for encouraging this work and for valuable discussions, and Joseph Maciejko, Frank Marsiglio, Min Ju Park, Myung-Joong Hwang, Alexander Hickey, and Sourav Biswas for helpful discussions. We gratefully acknowledge support from Quantum Horizons Alberta.

\appendix

\section{Equations of motion for the first moments}
From the Lindblad master equation, the equations of motion for the first moments are obtained as
\begin{equation}
    \frac{d\langle \mathcal{O}\rangle}{dt} = \mathrm{Tr}[\mathcal{O}\mathcal{L}[\rho]],
\end{equation}
where $\mathcal{O}=a_i,\: J^j$ with $i=1,2$ and $j=x,y,z$. Defining $\alpha_i=\langle a_i\rangle$ and $S^j=\langle J^j\rangle$, we obtain
\begin{equation}
\begin{aligned}
\frac{d\alpha_{1}}{dt} &= -(\kappa-\delta\kappa+i\Delta)\alpha_{1}-i\frac{2g}{\sqrt{N}}S^{x},\\
\frac{d\alpha_{2}}{dt} &= -(\kappa+\delta\kappa-i\Delta)\alpha_{2}+\frac{2g}{\sqrt{N}}S^{x},\\
\frac{dS^{x}}{dt} &= -\omega S^{y},\\
\frac{dS^{y}}{dt} &= \omega S^{x}-\frac{4g}{\sqrt{N}}S^{z}(\mathrm{Re}[\alpha_{1}]+\mathrm{Im}[\alpha_{2}]),\\
\frac{dS^{z}}{dt} &= \frac{4g}{\sqrt{N}}S^{y}(\mathrm{Re}[\alpha_{1}]+\mathrm{Im}[\alpha_{2}]).
\end{aligned}
\end{equation}

The steady states are obtained by setting the left-hand sides to zero. Solving these five equations yields two regimes: the trivial (normal) phase without photon condensation and the nontrivial (superradiant) phase with finite photon fields. The explicit solutions are presented in the main text.

\section{Gauge choice of the complex phase}
Within the effective model considered here, a phase transformation of the photon fields can be regarded as a gauge choice in our theory, $a_{j}\rightarrow e^{i\phi_{j}}a_{j}$. This gauge choice refers to a redefinition of the cavity operators within the effective theory and should be distinguished from an electromagnetic gauge transformation of a microscopic light-matter Hamiltonian. The dissipators, corresponding to single-photon losses, are invariant under this transformation, $\mathcal{D}[a_{j}]=\mathcal{D}[e^{i\phi_{j}}a_{j}]$, whereas the explicit form of the coherent part changes. For example, choosing $\phi_{1}=0$ and $\phi_{2}=-\pi/2$ leads to the Hamiltonian
\begin{equation}
    H= \Delta (a^{\dagger}_{1}a_{1}-a^{\dagger}_{2}a_{2}) + \frac{2\sqrt{2}g}{\sqrt{N}}J^{x}(x_{1}+x_{2}) + \omega J^{z}.
\end{equation}
In this gauge, the photon condensation in the superradiant phase is
\begin{equation}
    \alpha_{j} = \frac{2gS^{x}}{\sqrt{N}} \frac{(-1)^{j}\Delta -i(\kappa-(-1)^{j}\delta\kappa)}{\Delta^{2}+\kappa^{2}-\delta\kappa^{2}+2i\Delta\delta\kappa},
\end{equation}
for $j=1,2$. Physical predictions within the effective model are unchanged by this redefinition of the cavity operators.

\section{Expansion of the master equation}To describe Gaussian fluctuations, we expand the master equation around the mean-field solution. 
For each cavity mode $(j=1,2)$ we introduce fluctuation operators $d_j = a_j - \alpha_j$, where $\alpha_j=\langle a_j\rangle$ denotes the mean-field photon amplitude. The corresponding quadratures are $\delta x_j = x_j - \sqrt{2}\,\mathrm{Re}[\alpha_j]$ and $\delta p_j = p_j - \sqrt{2}\,\mathrm{Im}[\alpha_j]$.

At the mean-field level, the collective spin is rotated about the $y$ axis by an angle $\theta$, such that $S^{x} = \frac{N}{2}\sin\theta$ and $S^{z} = -\frac{N}{2}\cos\theta$. In the rotated frame aligned with the mean-field spin, spin fluctuations are described as bosonic excitations (magnons) via the Holstein-Primakoff transformation~\cite{holstein1940field},
\begin{equation}
\begin{pmatrix}
J^{x} \\ J^{y} \\ J^{z}
\end{pmatrix}
=
\begin{pmatrix}
\cos\theta & 0 & -\sin\theta \\
0 & 1 & 0 \\
\sin\theta & 0 & \cos\theta
\end{pmatrix}
\begin{pmatrix}
\sqrt{\frac{N}{2}}\, x_b \\
\sqrt{\frac{N}{2}}\, p_b \\
z_b
\end{pmatrix},
\end{equation}
where $b$ is the bosonic annihilation operator,
\begin{equation}
z_b = -\frac{N}{2} + b^\dagger b ,
\end{equation}
and
\begin{equation}
x_b=\frac{b+b^\dagger}{\sqrt{2}}, \:
p_b=\frac{b-b^\dagger}{i\sqrt{2}}
\end{equation}
are the quadratures of the magnon mode.

Expanding the Liouvillian to quadratic order yields
\begin{equation}
\tilde{\mathcal{L}}[\rho]
=
-i[\tilde{H},\rho]
+(\kappa-\delta\kappa)\mathcal{D}[d_1]
+(\kappa+\delta\kappa)\mathcal{D}[d_2],
\end{equation}
with the effective Hamiltonian
\begin{equation}
\tilde{H}
=
\Delta(d^{\dagger}_{1}d_{1}-d^{\dagger}_{2}d_{2})
+\Omega\, b^{\dagger}b
+\sqrt{2}\,\tilde{g}(b+b^{\dagger})(\delta x_{1}+\delta p_{2}),
\end{equation}
where $\Omega=\omega\cos\theta-(4g\sin\theta/\sqrt{N})(\mathrm{Re}[\alpha_{1}]+\mathrm{Im}[\alpha_{2}])$, $\tilde g=g\cos\theta$, and 
$\kappa_{\pm}=\kappa\pm\delta\kappa$. Constant and higher-order terms are omitted. This quadratic theory describes fluctuations around the mean-field solution and is valid in both the normal and superradiant phases in the large-$N$ Gaussian limit.

\section{Third-order exceptional point}\label{Appx_EP3}
We consider the time evolution of the wavefunction at the EP-3. The Hamiltonian can be transformed into a $3\times3$ Jordan block as
\begin{equation}
    \mathbf{S}_{3}^{-1}H_{\rm EP3}\mathbf{S}_{3}
    =\lambda_{0}\mathbb{I}_{3}+\mathbf{N}_{3},
\end{equation}
where $\mathbb{I}_{3}$ is the $3\times 3$ identity,
\begin{equation}
    \mathbf{N}_{3}=
    \begin{pmatrix}
        0 & 1 & 0 \\
        0 & 0 & 1 \\
        0 & 0 & 0
    \end{pmatrix},
\end{equation}
and $\mathbf{N}_{3}^{3}=0$. Then, the time-evolution operator is given by
\begin{equation}
    e^{-iH_{\rm EP3}t}
    =\mathbf{S}_{3}e^{-i\lambda_{0}t}
    \left(\mathbb{I}_{3}-i\mathbf{N}_{3}t-\frac{1}{2}\mathbf{N}_{3}^{2}t^{2}\right)
    \mathbf{S}_{3}^{-1}.
\end{equation}
Therefore,
\begin{equation}
    \psi(t)
    =\left(\mathbb{I}_{3}+\mathbf{C}_{1}t+\mathbf{C}_{2}t^{2}\right)
    e^{-i\lambda_{0}t}\psi(0),
\end{equation}
where $\mathbf{C}_{1}=-i\mathbf{S}_{3}\mathbf{N}_{3}\mathbf{S}_{3}^{-1}$ and $\mathbf{C}_{2}=\mathbf{C}_{1}^{2}/2$. The quadratic term in time is a characteristic consequence of the third-order Jordan block.

We extend our previous discussion of the quadrature variance obtained from the Langevin equation to the case of an EP-3. In Eq.~\eqref{Eq_V_1}, we assume $\mathbf{A}=\mathbf{S}'\mathbf{J}'\mathbf{S}'^{-1}$, where
\begin{equation}
    \mathbf{J}'
    =\mathrm{diag}(\lambda_{1},\lambda_{2},\lambda_{3},\ldots,\lambda_{n})
    +\mathbf{E}'_{12}+\mathbf{E}'_{23}.
\end{equation}
Here, $\mathbf{E}'_{ij}$ is an $n\times n$ matrix whose $(i,j)$ component is unity, while all other components vanish. At the EP-3, the three eigenvalues coalesce as $\lambda_{1}=\lambda_{2}=\lambda_{3}$. One can then show that
\begin{equation}
    e^{\mathbf{A}u}
    =\mathbf{S}'
    \left(\mathbf{\Omega}'_{\lambda}
    +f_{12}\mathbf{E}'_{12}
    +f_{23}\mathbf{E}'_{23}
    +f_{123}\mathbf{E}'_{13}\right)
    \mathbf{S}'^{-1},
\end{equation}
where $\mathbf{\Omega}'_{\lambda}=\mathrm{diag}(e^{\lambda_{1}u},\dots,e^{\lambda_{n}u})$, $f_{ij}=(e^{\lambda_{i}u}-e^{\lambda_{j}u})/(\lambda_{i}-\lambda_{j})$, and
\begin{equation}
    f_{123}
    =\sum_{j=1}^{3}
    \frac{e^{\lambda_{j}u}}
    {\prod_{k\neq j}(\lambda_{j}-\lambda_{k})}.
\end{equation}
When substituted into the Lyapunov equation [Eq.~\eqref{Eq_V_1}], the $f_{123}\mathbf{E}'_{13}$ term could generate stronger singular contributions to the covariance matrix than those at a second-order EP. Its quadratic dependence on the eigenvalue differences suggests a higher critical exponent, whose quantitative determination is left for future work.

\bibliography{BibRef}

\end{document}


\title{Supplemental Material for ``Enhanced dissipative criticality at an exceptional point"}

\author{Jongjun M. Lee}
\thanks{Contact author: jongjun@ualberta.ca}
\affiliation{Department of Physics, University of Alberta, Edmonton, Alberta T6G 2E1, Canada}
\affiliation{Quantum Horizons Alberta \& Theoretical Physics Institute, University of Alberta, Edmonton, Alberta T6G 2E1, Canada}

\date{\today}
\maketitle
\tableofcontents
\appendix

\section{Liouvillian spectrum and decay rate}
Expanding the Lindblad master equation around the mean-field solution, we obtain an effective quadratic master equation for the cavity photons and the magnon mode. Under the EP-tuned condition, the linearized dynamics is governed by the drift matrix
\begin{equation}
\begin{aligned}
\mathbf{A}=
\begin{pmatrix}
-\kappa_{-} & \Delta & 0 & 0 & 0 & 0 \\
-\Delta & -\kappa_{-} & 0 & 0 & -2g & 0 \\
0 & 0 & -\kappa_{+} & -\Delta & +2g & 0 \\
0 & 0 & \Delta & -\kappa_{+} & 0 & 0 \\
0 & 0 & 0 & 0 & 0 & \omega \\
-2g & 0 & 0 & -2g & -\omega & 0
\end{pmatrix},
\end{aligned}
\end{equation}
in the normal phase where $\kappa_{\pm}=\kappa\pm\delta\kappa$. At the critical coupling $g=g_c$, this matrix satisfies $\text{Det}[\mathbf{A}]=0$. The critical coupling is given by
\begin{equation}
g_{c} = \frac{1}{4} \sqrt{\frac{\omega D}{\Delta\kappa \delta \kappa}},
\end{equation}
with
\begin{equation}
D=4\Delta^{2}\delta\kappa^{2}+(\Delta^{2}+\kappa^{2}-\delta\kappa^{2})^{2}.
\end{equation}
Near the critical point, the slowest Liouvillian eigenvalues are small. Keeping the leading orders in $\lambda$, the characteristic equation reduces to
\begin{equation}
\text{Det}[\lambda \mathcal{I}-\mathbf{A}]\simeq A(\kappa+\lambda)+B\lambda^{2}=0,
\end{equation}
where
\begin{equation}
\begin{aligned}
A\simeq &\,32 g_{c}\omega \delta\kappa 
\sqrt{\kappa^{2}-\delta\kappa^{2}+2\sqrt{\kappa^{2}(\kappa^{2}-\delta\kappa^{2})}}
\,(g_{c}-g),\\
B=&\,4\Big[(2\kappa^{2}+\omega^{2})
(\kappa^{2}-\delta\kappa^{2}+\sqrt{\kappa^{2}(\kappa^{2}-\delta\kappa^{2})})
+\kappa^{2}\omega^{2}\Big].
\end{aligned}
\end{equation}
Since $A\ll B$ near the critical point and $A,B>0$, the eigenvalues are approximately
\begin{equation}
\lambda\simeq \frac{1}{2B}\left(-A\pm 2i\sqrt{\kappa B}\sqrt{A}\right).
\end{equation}
Therefore the real and imaginary parts scale as
\begin{equation}
\mathrm{Re}[\lambda]\propto -(g_c-g), \quad 
\mathrm{Im}[\lambda]\propto \pm\sqrt{g_c-g}.
\label{Eq_Re_Im_1}
\end{equation}

\begin{figure}
    \centering
    \includegraphics[width=0.75\linewidth]{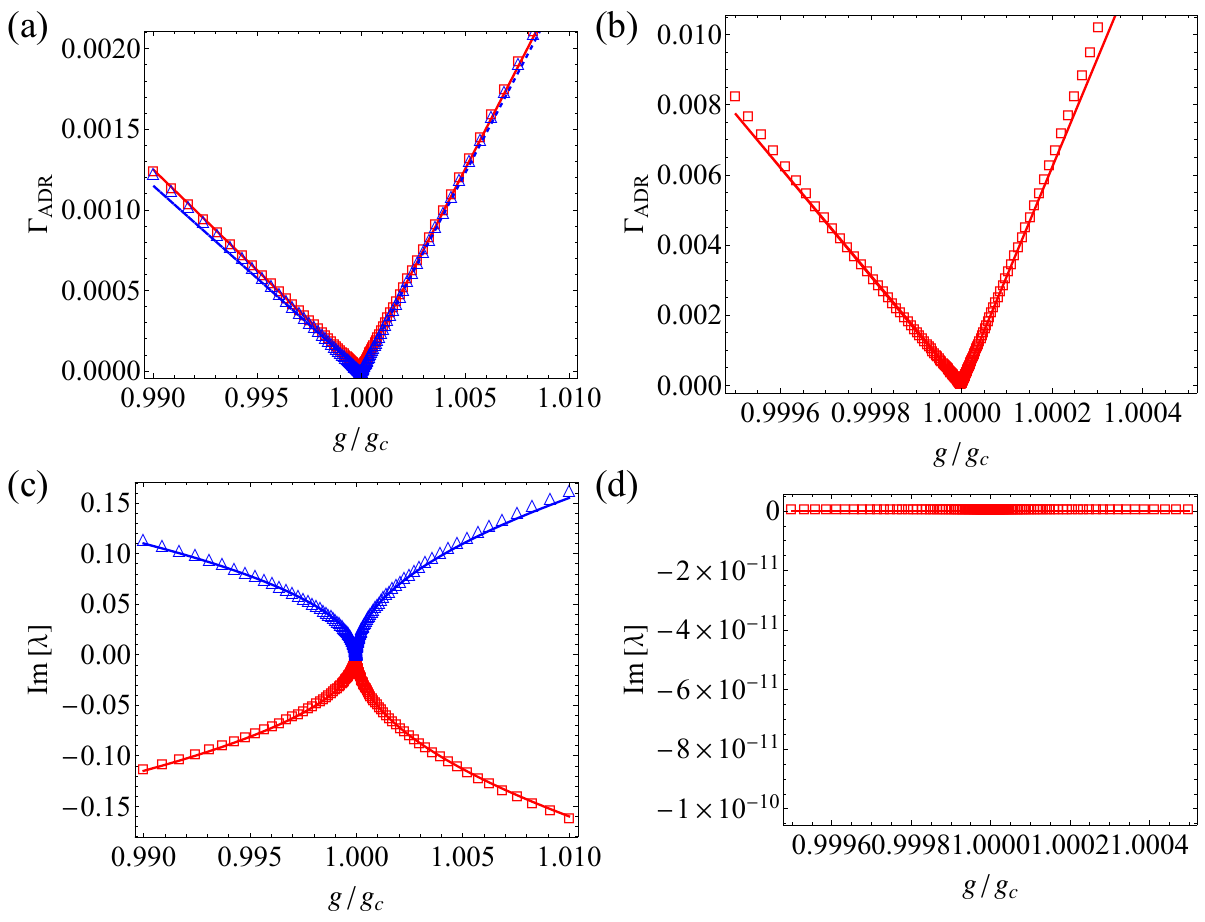}
    \caption{Liouvillian spectrum near the critical point $g_c$. (a,b) Asymptotic decay rate $\Gamma_{\rm ADR}$ as a function of the coupling strength $g$ around $g_c$. (c,d) Imaginary part of the Liouvillian eigenvalues $\mathrm{Im}[\lambda]$ for the slow modes. Panels (a,c) correspond to the EP-tuned condition, where two slowest modes appear, while (b,d) correspond to the off-tuned case with a single slowest mode. Symbols denote numerical data and solid lines indicate fits to $|g-g_c|$ for $\Gamma_{\rm ADR}$ and $\pm\sqrt{|g-g_c|}$ for $\mathrm{Im}[\lambda]$.}
    \label{SFig1}
\end{figure}

In Fig.~\ref{SFig1}, we plot the Liouvillian spectrum of the slowest modes as a function of the coupling strength $g$ near the critical point. Both the real and imaginary parts approach zero at criticality and follow the scaling behavior in Eq.~(\ref{Eq_Re_Im_1}) in both the normal and superradiant phases.

\section{Divergence of the quadrature variance}
The linearized Langevin equation leads to the following Lyapunov equation for the steady-state covariance matrix $\mathbf{V}$~\cite{mari2012cooling,woolley2014two}.
\begin{equation}
    \mathbf{A}\mathbf{V}+\mathbf{V}\mathbf{A}^{\rm T} + \mathbf{D} = 0,
\end{equation}
where $\mathbf{A}$ and $\mathbf{D}$ denote the drift and diffusion matrices, respectively. 
The covariance matrix is defined in terms of symmetrized correlations of the quadrature operators as
\begin{equation}
\mathbf{V} = \begin{pmatrix}
    \langle \delta x_{1}\delta x_{1}\rangle_{s} & \langle \delta x_{1}\delta p_{1}\rangle_{s} & \langle \delta x_{1}\delta x_{2}\rangle_{s}  & \langle \delta x_{1}\delta p_{2}\rangle_{s}  & \langle \delta x_{1}\delta x_{b}\rangle_{s}  & \langle \delta x_{1} p_{b}\rangle_{s} \\
    \langle \delta p_{1}\delta x_{1}\rangle_{s} & \langle \delta p_{1}\delta p_{1}\rangle_{s} & \langle \delta p_{1}\delta x_{2}\rangle_{s}  & \langle \delta p_{1}\delta p_{2}\rangle_{s}  & \langle \delta p_{1}\delta x_{b}\rangle_{s}  & \langle \delta p_{1} p_{b}\rangle_{s} \\
    \langle \delta x_{2}\delta x_{1}\rangle_{s} & \langle \delta x_{2}\delta p_{1}\rangle_{s} & \langle \delta x_{2}\delta x_{2}\rangle_{s}  & \langle \delta x_{2}\delta p_{2}\rangle_{s}  & \langle \delta x_{2}\delta x_{b}\rangle_{s}  & \langle \delta x_{2} p_{b}\rangle_{s} \\
    \langle \delta p_{2}\delta x_{1}\rangle_{s} & \langle \delta p_{2}\delta p_{1}\rangle_{s} & \langle \delta p_{2}\delta x_{2}\rangle_{s}  & \langle \delta p_{2}\delta p_{2}\rangle_{s}  & \langle \delta p_{2}\delta x_{b}\rangle_{s}  & \langle \delta p_{2} p_{b}\rangle_{s} \\
    \langle x_{b}\delta x_{1}\rangle_{s} & \langle x_{b}\delta p_{1}\rangle_{s} & \langle x_{b}\delta x_{2}\rangle_{s}  & \langle x_{b} \delta p_{2}\rangle_{s}  & \langle x_{b} \delta x_{b}\rangle_{s}  & \langle x_{b} p_{b}\rangle_{s} \\
    \langle p_{b} \delta x_{1}\rangle_{s} & \langle p_{b}\delta p_{1}\rangle_{s} & \langle p_{b} \delta x_{2}\rangle_{s}  & \langle p_{b}\delta p_{2}\rangle_{s}  & \langle p_{b}\delta x_{b}\rangle_{s}  & \langle p_{b} p_{b}\rangle_{s},
\end{pmatrix}
\end{equation}
where $\langle o_{1}o_{2}\rangle_{s} = \frac{1}{2}(\langle o_{1}o_{2}\rangle+ \langle o_{2}o_{1}\rangle)$. The quadrature operators are defined as
\begin{equation}
    \delta x_{j} = \frac{\delta a_{j}+\delta a^{\dagger}_{j}}{\sqrt{2}},\:
    \delta p_{j} = \frac{\delta a_{j}-\delta a^{\dagger}_{j}}{i\sqrt{2}},\:
    x_{b} = \frac{b+b^{\dagger}}{\sqrt{2}},\:
    p_{b} = \frac{b-b^{\dagger}}{i\sqrt{2}},
\end{equation}
with $\delta a_{j} = a_{j}-\langle a_{j}\rangle$ for $j=1,2$.

We numerically compute the covariance matrix $\mathbf{V}$ as a function of the coupling strength $g$ and extract the critical exponents via power-law fitting. While the main text focuses on the number fluctuation and the purity, here we present representative results for several quadrature components. The results are summarized in Table~\ref{Table_summary_2}.

We find that the exceptional point enhances the divergence of most components. The momentum quadrature fluctuation of the magnon exhibits a critical exponent $-1$, in contrast to the exponent $-2$ found for other quantities. In the absence of the exceptional-point condition, this quantity remains constant near the critical point (exponent $0$), indicating a clear enhancement. The component $\langle x_{b}p_{b}\rangle_{s}$ remains zero near the critical point both with and without tuning to the exceptional point.

\begin{table}[t!]
\setlength{\tabcolsep}{6pt}
\begin{tabular}{lccccccccc}
\hline \hline 
& $\langle \delta x^{2}_{1}\rangle_{s}$ & $\langle \delta p^{2}_{1}\rangle_{s}$ & $\langle \delta x_{1}\delta p_{1}\rangle_{s}$ & $\langle \delta x^{2}_{2}\rangle_{s}$ & $\langle \delta p^{2}_{2}\rangle_{s}$ & $\langle \delta x_{2}\delta p_{2}\rangle_{s}$ & $\langle x^{2}_{b}\rangle_{s}$ & $\langle p^{2}_{b}\rangle_{s}$ & $\langle x_{b} p_{b}\rangle_{s}$ \\ \hline
Exceptional & $-2$ & $-2$ & $-2$ & $-2$ & $-2$ & $-2$ & $-2$ & $-1$ & - \\ \hline
Non-exceptional & $-1$ & $-1$ & $-1$ & $-1$ & $-1$ & $-1$ & $-1$ & $0$ & - \\ \hline
\end{tabular}
\caption{Summary of the critical exponents defined by $X \propto |g - g_{c}|^{\nu_X}$. 
``Exceptional” and ``Non-exceptional” refer to the extended model at and away from the exceptional-point condition, respectively. $\langle \delta x^{2}_{1,2} \rangle_{s}$, $\langle \delta p^{2}_{1,2} \rangle_{s}$, and $\langle \delta x_{1,2}\delta p_{1,2} \rangle_{s}$ denote the fluctuations of the quadrature variances for cavity modes 1 and 2, respectively. $\langle x^{2}_{b} \rangle_{s}$, $\langle p^{2}_{b} \rangle_{s}$ and $\langle x_{b}p_{b} \rangle_{s}$ denote the fluctuations of the magnon quadrature variances.}
\label{Table_summary_2}
\end{table}

\bibliography{BibRef}